\documentstyle[a4,12pt,epsfig,graphicx]{article}
\begin{document}
\titlepage
\begin{flushright}
SPhT 98/058\\
hep-th/9806176 \\
June 1998
\end{flushright}
\vskip 1cm
\begin{center}
{\bf \Large
Limits of Matrix Theory in Curved Space}
\end{center}

\vskip 1cm
\begin{center}
{\large Ph. Brax$^{a}$\footnote{email: brax@spht.saclay.cea.fr} 
$\&$ T. Wynter$^{b}$\footnote{email: wynter@spht.saclay.cea.fr}}
\end{center}
\vskip 0.5cm
\begin{center}
$^a$ {\it Service de Physique Th\'eorique, 
CEA-Saclay\\
F-91191 Gif/Yvette Cedex, France}\\
\end{center}
\vskip .5 cm
\begin{center}
\end{center}
\vskip 2cm
\begin{center}
{\large Abstract}
\end{center}
\vskip .3in \baselineskip10pt{We study curved space versions of matrix
string theory taking as a definition of the theory a gauged matrix
sigma model. By analyzing the divergent 
terms in the  loop expansion for the 
effective action we reduce the problem to a simple matrix generalization of
the standard string theory beta function calculation. It is then
demonstrated that the model can only be consistent for Ricci flat
manifolds with vanishing six-dimensional Euler density.}
\bigskip
\rightline{SPhT-98/}

\noindent
\newpage
\baselineskip=1.5\baselineskip
\section{Introduction}

Matrix string theory \cite{motl} \cite{bs} \cite{dvsq} is hoped to
 provide a non-perturbative definition of type IIA string theory.
The free string states have
a clear interpretation in this picture. They are formed from ``winding
sectors'' in which large
numbers of eigenvalues form, via twisted boundary conditions, 
long string configurations composed of an order $N$ number of 
eigenvalues. The large $N$ limit corresponds to taking a finer and
finer discretization of the light-cone string world-sheet into
infinitesimal strips, and corresponds to taking the conformal limit
of the theory. From this conformal field theory point of view 
string interactions can be argued to be described by 
an irrelevant, local CFT operator \cite {dvsq}. Dimensional 
analysis then requires  the operator to be associated with a linear 
power of the string
coupling constant $g_s$. This  analysis however leaves unclear what power 
of $N$, if any is associated with the interaction. 
Progress has been made in constructing classical solutions corresponding to 
string interactions \cite{wyn1}\cite{GHV}\cite{BBN} 
but it is not yet known how to calculate the quantum 
fluctuations to confirm the 
string interaction weight directly from the SYM theory.

In string theory the classical equations
of motion for the effective theory can be found either from tree level
string scattering amplitudes or from the consistency conditions
(conformal invariance) of the action for a non-interacting string
propagating in a curved space with background fields. It is thus worth
exploring what can be learnt from the consistency conditions on curved
space versions of matrix string theory. In 
\cite{mikes}\cite{miked}\cite{DOK} gauged versions of matrix sigma 
models were proposed as a
description for D-brane actions in curved space at finite N, and
it was suggested that they might also be used as descriptions of matrix
theory in curved space.  These
actions were built from a simple set of axioms, the most important
being that they have a $U(N)$ invariance are built from a single trace
and for diagonal matrices describe $N$ identical copies of the
standard string sigma model.
We will describe the matrix string model coupled to a curved
background using a slight modification of these axioms. 

We study here the conditions imposed on the curved space versions
of matrix string theory  in which perturbative
string theory is hoped to be recovered. The
calculations resemble closely the determination of the string beta 
function, evaluated using the background field method. This was
already observed in 
\cite {mikes}.
 We will show below that such
models are only consistent for an extremely limited class of
manifolds i.e. Ricci flat manifolds with vanishing Euler class.
An example of which is provided by the direct product $\bf
M=S\times C$ where $\bf S$ is a hyper-K\"ahler surface. This is for
instance the case of the ALE spaces\cite{dougmoo}.

The calculation is performed in the context of matrix string
theory, but this  result applies also to the quantum 
mechanics of D0 branes.

We begin in section 2. by recalling the essential elements of matrix
string theory.  In section
3. we review the proposals for D-brane actions/ matrix theory in
curved space put forward by Douglas et. al. 
\cite{mikes}\cite{miked}\cite{DOK}. 
In section 4 we focus on the divergent part of calculation of the
effective action  and show that it 
reduces to a simple
matrix generalization of the string theory beta function calculation. 
Having mapped the 
calculation to that of the string beta function we use known 4-loop 
results \cite {gris}
to demonstrate that the effective action can only be consistent for 
Ricci flat manifolds with vanishing six-dimensional Euler
density.

\section{$N=8$ Two Dimensional Super Yang-Mills and Matrix String Theory}
In this section we summarize the essential ingredients of the
correspondence between the dimensional reduction of ten dimensional
super Yang-Mills theory and type IIA string theory.  The two
dimensional action reduces to
\begin{equation}
S=\int d\tau d\sigma \hbox {Tr} [ {1\over 2} (D_{\alpha}X^{I})^2 
+{i\over 2}\Theta^T\gamma^{\alpha}D_{\alpha}\Theta 
-{1\over 4}F^2_{\alpha\beta}+{1\over 4g_s^2}[X^I,X^J]^2
+{1\over 2 g_s}\Theta^T\gamma_I[X^I,\Theta]]
\label {SYM}
\end {equation}
The fields are $N\times N$ Hermitean matrices. The index $I$ runs 
from 1 to 8 and the sixteen fermions split into the $8_s$ and $8_c$ 
spinorial representations of $SO(8)$. The string coupling constant 
of the type IIA string theory is $g_s$. The coordinate $\sigma$ 
lives between 0 and 2$\pi$.

According to \cite {dvsq} the weakly coupled string is to be obtained from 
the $g_s\to 0$ limit corresponding to the infra-red limit of the 
SYM theory. In this regime the matrices commute and describe 
 strings in the light cone frame. The corresponding 
action evaluated for these configurations is the sum of $N$ 
replicas of the light cone Green-Schwarz 
action.  In this limit the matrix coordinates can always be 
diagonalized using unitary transformations $U$
\begin{equation}
X^I=Ux^IU.
\end{equation}
The matrix $U$ is defined up to an element $g$ of the Weyl group of $U(N)$
permuting the eigenvalues. 
\begin{equation}
U(\sigma +2\pi)=U(\sigma)g, \ \ x^I(\sigma+2\pi)=gx^I(\sigma)g^{\dagger }
\label{U}
\end{equation}
The infra-red regime is then identified with the two-dimensional 
conformal field theory described by the $N=8$ sigma model on the 
target space 
\begin{equation}
S^N R^8=(R^8)^N/S_N
\end{equation}
The freely propagating strings in the light cone frame are identified 
in the limit $N\to \infty$ with the cycles of the eigenvalues $x^I$ 
under the permutation group.

It is useful at this point to reformulate the action (\ref{SYM}) in a
way that generalizes easily to a curved background.
A supersymmetric and gauge
invariant action can be written using the $d=4$, $N=1$ superfield
formalism. The four gauge fields belong to a vector multiplet $V$
while the bosonic fields belong to three chiral multiplets $\Phi^i$.
The eight bosonic fields  of the original action thus split into a
group of six belonging to the three chiral multiplets and two obtained
by dimensional reduction of the 4d gauge fields belonging to the
vector multiplet. This formulation breaks the global $SO(8)$ symmetry
into $SO(6)\times SO(2)$.  The full $SO(8)$ symmetry is restored by
going to the Wess-Zumino gauge. The Lagrangian is
\begin{equation}
S={1\over \alpha '}\hbox {tr} 
(\int d^2x d^4\theta e^{gV}\Phi e^{-gV}\bar\Phi^{\dagger}
+{1\over 64g^2}\int d^2x d^4\theta W^2 +
{ {ig}\over 3!\sqrt{\alpha '}}
\int d^2xd^2\theta \epsilon_{ijk}\Phi^i\Phi^j\Phi^k+cc)
\label{SYM14}
\end{equation}
where $g^{-2}=\alpha ' g_s^2$ is the YM coupling constant and 
$W_{\alpha}=\bar D^2 e^{gV}D_{\alpha}e^{-gV}$.
The two derivative Lagrangian of the 4d $N=4$ SYM theory is finite, 
in particular the beta function vanishes. 

In this article we will be focusing uniquely on the two derivative part of
the effective action. To set up the formalism for later use let us
briefly describe how the calculation proceeds in terms of
superfields. This is just a rerun of how the background 
superfield formalism can be used to perturbatively 
show the finiteness of $d=4$,
$N=4$ SYM. We send the reader to \cite{GSR}
for fuller details. The
background superfields in our case will be the 
diagonal field configurations corresponding to long strings.
           
One 
is interested in separating the superfields into a background 
 configuration and the quantum fluctuating parts. The 
fluctuating parts comprise terms inside and outside the Cartan 
subalgebra. It is most 
convenient to use the background field formalism 
\cite{GSR}\cite{onethou}\cite{west}
a review of which can be found in the appendix.
We decompose the vector superfields according to 
\begin{equation}
e^{V_T}=e^{V_B\over 2 }  e^{V}e^{V_B\over 2}
\label{VBv}
\end{equation}
where $V_T$ is the total superfield while $V_B$ is the background 
configuration and $V$ the fluctuating part.  
The background chiral superfields can be split into Cartan 
$\Phi_h$ and fluctuating quantum parts $\phi_h,\ \phi$
\begin{equation}
\Phi=\Phi_h+\phi_h +\phi
\label{bacfluc}
\end{equation}
where $\phi$ does not belong to the Cartan subalgebra.
 The renormalized Lagrangian 
is obtained after integrating over the fluctuating parts.
This integration is nothing but the renormalization process of the 
$N=4$ SYM theory reduced to two dimensions. It is well known 
\cite {onethou} that  the two derivative renormalized action is 
finite.
The result of the integration over the fluctuating parts gives 
the original action when the fields are given by their 
background values. This implies that the effective action is 
\begin{equation}
S_{eff}={1\over \alpha'}( \int d^2xd^4\theta \bar\Phi_d^{\dagger} 
\Phi_d +  {1\over 64} \int d^2x d^2\theta W^2)
\end {equation}
where $W_{\alpha }= \bar D^2 D_{\alpha } V_B$. Expanding in component
fields and putting to zero the F and D terms yields $N$ copies of the
flat space string theory action.  The crucial point of the present
derivation is the finiteness of the $N=4$ SYM theory. In the following
we will apply the same method to the matrix string theory in a curved
background.

\vskip 1 cm

\section {Curved Space Actions}

Candidate formulations for D-brane actions in curved space have been
been proposed in \cite{mikes}\cite{miked}\cite{DOK}. 
For small curvatures a single
D-brane is described by the Born-Infeld theory. The crucial point is
that this contains a $U(1)$ gauge field which becomes non-Abelian when
$N$ D-Branes coincide. In the low energy regime this reduces to a SYM
theory on the world-volume of the D-branes. In curved space the
D-brane action should combine the non-Abelian nature of the gauge
theory and a fraction of the original sixteen supersymmetries
preserved by the D-brane configuration.

A set of axioms have been proposed in \cite{miked}\cite{DOK} 
to describe the possible
actions.  A particularly natural set of D-brane actions in this
context are those obtained from the the dimensional
reduction of a 4d $(N=1)\ \ \ U(N)$ SYM theory to $d+1$
dimensions \cite{mikes}. The curved background is a 3d complex K\"ahler manifold
whose metric depends on a K\"ahler potential $K$. The vector superfields
contain $(3-d)$ real flat coordinates. Notice that the splitting of the
background manifold implies that the original $SO(8)$ global symmetry
is reduced to $SO(3-d)$. The case $d=1$ corresponds to the matrix
string theory while $d=0$ is a curved version of the matrix model for
M-theory.

In a setting adapted to our purposes the axioms amount to the
following four requirements for the D-brane action defined on a 3
dimensional K\"ahler manifold ${\cal M}$.

\vskip .5 cm
a) The classical moduli space, determined by the vanishing of the $D$
and $F$ terms of the SYM theory, is the symmetric product ${\cal M}^N/S_N$.
\vskip .5 cm
b) The generic unbroken gauge symmetry is $U(1)^N$.
\vskip .5 cm
c) Given non-coincident branes at points $p_i\ne p_j$, all 
states charged under $U(1)_i\times U(1)_j$ have mass 
$m_{ij}=d(p_i,p_j)$ the distance along the shortest geodesic 
between the two points.
\vskip .5 cm
d) The action is a single trace.

\vskip 1cm
\noindent These axioms imply  that the action in curved space reads
\begin{equation}
S={1\over\alpha '}\hbox {tr}(\int d^{d+1}xd^4\theta
K(e^{gV}\Phi e^{-gV},\bar\Phi ^{\dagger}) 
+(\int d^{d+1}x d^{4}\theta W(\Phi) 
+{1\over 64 g^2}\int d^{d+1}x d^2\theta  W^{\alpha}W_{\alpha} +cc))
\end{equation}
The analysis of axiom a) leads to the following form for the superpotential
\begin{equation}
W= \epsilon_{ijk} a^i(\Phi) [\Phi^j,\Phi^k]
\end{equation}
where $a^i(\Phi)$ is a holomorphic vector field in the adjoint
representation of the gauge group.  In the following we will choose
such a superpotential but consider a less restrictive set of
axioms. Effectively we will relax axiom c) and use the most general
K\"ahler potential allowed by supersymmetry and gauge invariance.  
We will use the fact that there exists around each point of the
moduli space a set of normal K\"ahler coordinates. These coordinates are
such that locally
\begin{equation}
K(z,\bar z)= z\bar z + \sum {1\over {L_R^{n-2}}} 
K_{I_1..I_p\bar I_{p+1}..\bar I_n} 
z^{I_1}..z^{I_p}\bar z^{I_{p+1}}..\bar z^n
\end{equation}
The existence of this expansion is guaranteed up to an analytic change
of coordinates on the curved manifold. 
By definition the $K_{I_1..I_p\bar I_{p+1}..\bar I_n}$ are symmetric
with respect to arbitrary reorderings of the holomorphic indices and
arbitrary reorderings of the antiholomorphic indices.
Finally since we are dealing
with matrices there is a question of ordering in the K\"ahler
potential. The most natural ansatz is to assume that all terms in
the K\"ahler potential are symmetrized products of matrices, but there
could be more general orderings. The fourth order term for example can
be written as
\begin{equation}
K_{I K \bar J \bar L} \bigl[
\delta\,\Phi^I \Phi^K \bar\Phi^{\bar J} \bar \Phi ^{\bar L}\,+\,
\tau\,\Phi^I \bar\Phi^{\bar J} \Phi^K \bar \Phi ^{\bar L} \bigr],
\end{equation}
where $\delta$ and $\tau=1-\delta$ are constants. 
It is also possible for the K\"ahler potential to contain terms
proportional to commutators of matrices since these vanish for the
classical moduli space (diagonal matrices).
In fact it was found
in \cite{DOK} that imposing the axioms stated above constrains the
fourth order term to be the totally symmetrized product ($\delta=2/3$,
$\tau=1/3$) with no additional terms corresponding to commutators. 

Let us first use a very naive argument to
justify the link between the matrix string theory on a curved
background and the type IIA string theory in curved
space. Substituting the diagonal matrices describing the moduli space
$\cal M$ in the action leads to a sum of $N$ copies of the $U(1)$
gauged sigma model in 2d. The gauge part of the action describing the
flat component of the background manifold decouples and one is left
with $N$ copies of the sigma model defined by the background curved
manifold
\begin{equation}
{1\over\alpha '} \int d^2x d^4\theta K(\Phi,\bar \Phi)
\end{equation}
where $\Phi$ represents one of the $N$ components. The analysis of
this action reveals that there are two dimensional UV
divergences. These logarithmic divergences can be cancelled up to three
loop order by imposing that the Ricci tensor vanishes
\begin{equation}
R_{I\bar J}=0
\end{equation}
This is the usual Einstein equation as deduced from the conformal
invariance of string theory.  At four loop order this is not true
anymore, the beta function is non-zero for Ricci-flat manifolds. The
divergence is proportional to
\begin{equation}
R_{hkmn}R^{h}_{rs}\ ^{n}(R^{ksrm}+R^{kmrs})
\end{equation}
when expressed in terms of the underlying real coordinates.
This is equivalent to the result obtained from the calculation of the
 four-graviton scattering for type IIA theory.  This leads to a
 correction of the effective 10d supergravity action and the familiar
 $R^4$ term.

It seems therefore that a naive application of matrix theory in curved
space leads to the correct identification of the string
equations. This is misleading as a detailed analysis expounded in the
following will show.

\vskip 1 cm

\section {The Effective Action in  a Curved Background}

In the previous section we have defined the curved background version
of the matrix string theory.  This involves an explicit splitting
between the six curved coordinates represented by a non-linear sigma
model coupled to $SU(N)$ YM fields and the two coordinates obtained by
dimensional reduction of the four dimensional YM gauge fields.  We are
interested in the equivalence between this theory and string theory in
a curved background. In particular we have seen that a naive
calculation of the effective action for diagonal configurations leads
to the string equations.  In this section we reexamine this issue by
properly integrating over the background fluctuations to arrive at an
effective action for the diagonal configuration. 
We will focus solely on the divergent contributions to the K\"ahler
potential. We will show that the resulting effective action can only
be consistent for a very limited class of manifolds. 

\subsection{Superfield reduction}
 Separating the chiral superfields and the 
vector superfields into diagonal and off-diagonal parts 
the effective action for the diagonal fields is obtained by 
integrating over the off-diagonal elements $\phi$ and $v$ and 
the fluctuations of the diagonal parts $\phi_d$ and $v_d$. The 
resulting effective action possesses a modified K\"ahler 
potential $K_R$ in such a way that 
\begin{equation}
S_{eff}={1\over \alpha '} \int d^2x d^4\theta K_R 
(\Phi_d,\bar \Phi_d)
\end{equation}
The superpotential is not renormalized and vanishes for 
diagonal configurations.
The renormalized K\"ahler potential is  obtained after removing 
the UV divergences leading to poles in 
$1\over \epsilon$ when using dimensional regularization. These 
poles correspond to the logarithmic divergences of the sigma 
models in two dimensions. 
 
Let us see in more details how this is implemented in the $N=2$
two dimensional SYM context. It is well known that the flat space
action with simple quadratic K\"ahler potential (\ref{SYM14}) is finite
\cite{onethou}
so it is only diagrams containing higher order terms in the K\"ahler
potential that can lead to divergences. The simplest such term is
\begin{equation}
K_{I\bar J K\bar L} \Phi^I e^{-gv}\bar\Phi^{\bar J} e^{gv} 
\Phi^K e^{-gv}\bar \Phi ^{\bar L} e^{gv},
\label{Kterm}
\end{equation}
along with its symmetrized partners. Each field $\Phi$ can be split
into a background part and a fluctuation part (\ref{bacfluc}). The
fields $v$ (see equation(\ref{VBv})) are the fluctuating
part of the vector superfields and the
exponentials can then be expanded to arbitrary order. All fluctuation
fields are contracted, with the 
loop diagrams being easily deduced from the usual 
rules for the propagators and the vertices. Below we give the relevant
propagators and vertices, with their appropriate factors of ${\cal D}^2$
and $\bar{\cal D}^2$ (\ref{calD}). Since we are only interested in the 
divergent part of the Feynman diagram expansion we can simply 
replace the background covariant derivatives ${\cal D}^2$ (\ref{calD})
by ordinary fermionic derivatives $D^2$.
\begin{figure}[h]
\includegraphics{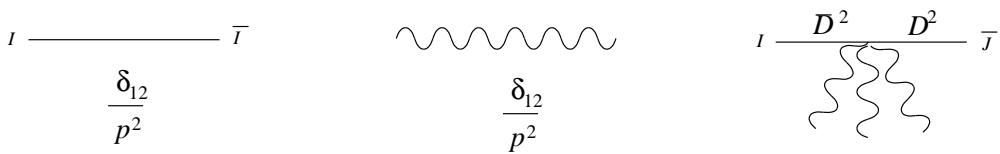}
\label{Fig1}\caption{Propagators and vertices}
\end{figure}
The diagonal background fields lead to the off-diagonal quantum fields
acquiring a mass. Again since we are only interested here in the 
ultraviolet divergences we ignore these masses. In this context we 
can also ignore the propagators between two chiral fields and those
between two antichiral fields. Likewise the contribution from the
superpotential vertices can be ignored as we will discuss below.
We are thus left with the Feynman vertices and propagators shown in
figure 1. These connect together the quantum fields in the expansion
of the K\"ahler potential such as (\ref{Kterm}). A typical K\"ahler
potential vertex is illustrated below. Only the quantum fields are 
represented. All chiral legs carry a factor $\bar{D}^2$ and all
anti-chiral legs a factor of $D^2$.
\begin{figure}[h]
\hskip 150pt\includegraphics{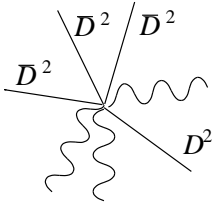}
\label{Fig2}\caption{Typical K\"ahler potential vertex}
\end{figure}

The simplest divergent contributions are those coming from connecting 
together chiral and antichiral fields, and with no vector superfields
participating. These contributions are matrix generalizations of the
divergent diagrams calculated in string theory in the determination of
the beta function by the background field method. We also, however,
have to consider the contribution of loops containing one or more
vector superfields and/or superpotential insertions.

Let us first consider the vector superfield contribution. A closed
vector superfield loop is automatically zero since it is proportional
to $\delta(0)$ in superspace. A single loop formed from a vector
propagator and a chiral propagator can be seen to be ultraviolet
finite.  It has one momentum integral, contributing $p^2$, four
propagators, contributing $p^{-4}$ and four $D$s. Dimensionally the
four $D$s are the equivalent of two momenta however they are used up
in the identity $\delta_{12}\bar{D}_1^2D_1^2\delta_{12}=16\delta_{12}$
in integrating over the $\theta$ variables (see for example
\cite{onethou}\cite{west}). The total diagram thus has divergence $-2$.

One can proceed systematically in this way to find the divergence for
a general loop diagram. If one has $L$ loops with $P$ propagators of
any type, $C$ of which are chiral-antichiral propagators, we have the
degree of divergence
\begin{equation}
{\rm div}= 2L - 2P +2C -2L=-2(P-C).
\label{div}
\end{equation}
The first $2L$ comes from the integral over the $2L$ loop momenta
and the $-2P$ from the propagators. We are only
considering the vertices shown in figures 1 and 2 so all
chiral-antichiral propagators come with four factors of $D$,
equivalent to two momenta and hence the contribution $2C$. 
The final $-2L$ in (\ref{div}) comes from the fact that for each loop
we need four $D$s to obtain a non-zero result when performing the
final $\theta$ integral connected with a loop. 
We thus see that for divergent diagrams $P=C$. 

Finally superpotential insertions always reduce the degree of
divergence since one of their internal legs has no factors of $D$.
We are thus left with examining the divergences due to chiral 
diagrams with no superpotential insertions and no gauge fields.

\subsection{Chiral diagrams}
We have thus reduced the calculation of the divergent part of the loop
expansion for the effective action to a matrix generalization of the
string theory beta function calculation. 

As stated above it is known that in string theory the one, two or
three loops divergent contributions disappear for Ricci flat manifolds
whilst at four loops there is a correction that only disappears for
manifolds with a vanishing six dimensional Euler density.
 The divergences  lead to the famous
$R^4$ term being added to the low energy effective action for the
massless modes of the string. In other words Ricci flatness is a low
order approximation corrected by terms of higher order in $\alpha'$.

For the curved space versions of matrix theory however there are two
types of chiral diagrams. Firstly there are those 
coming from the expansion of the K\"ahler potential in 
terms of the diagonal fluctuations only. This is nothing  but 
$N$ copies of the two dimensional sigma model with values in 
a six dimensional complex K\"ahlerian manifold. 
Secondly there are 
diagrams involving the off-diagonal part of $\phi$. These will lead to
divergent terms involving one or more diagonal elements i.e. to terms
consisting of products of traces.  Since these are not included in the
original action they have to be set to zero. In other words we
find that each loop
order has to be individually set to zero. This is a more stringent
restriction than in string theory. 

Retaining only at  each loop order the contribution due to the
diagonal matrices is a simple generalization of the string
theory beta function result (we study this question below) we see
that, in particular, the four loop term has to be set to zero. This
implies that the curved manifold must be Ricci flat with a
vanishing Euler class. This is for instance the case of products
$\bf M\times C$ where $M$ is hyper-K\"ahler. In particular the ALE
spaces are good candidates for a description of matrix string theory in
curved space.

 This result thus
restricts quite severely the range of applicability of matrix sigma
models as descriptions of matrix theory in curved space.
Up to now  we  have only considered the terms due to the diagonal
matrices. This is not sufficient to guarantee the finiteness of
the model. We now turn to off diagonal contributions. We will
only examine them at the one loop order.

\subsection{1 loop contribution}

It is not immediately obvious that Ricci flat metrics lead to
vanishing one, two and three loop contributions for the matrix sigma
model.  Indeed the contribution of the off-diagonal matrices needs
to be carefully examined. It is important to measure their relevance  
at least for the first
non-trivial term. Failure of the cancellation process at this level  
would almost certainly 
lead to the conclusion that matrix theory can only be consistent for
flat space. The first non-trivial test involves
the sixth order term in the expansion of the K\"ahler potential. 
We show that, by a particular choice of ordering and the addition of a
particular commutator term (that vanishes on the classical moduli
space), this contribution will disappear for Ricci flat metrics. The
condition to be satisfied for this to be the case is identical to one
of the mass conditions deduced in \cite{DOK}. 

Let us first focus on the 1-loop contribution. 
For $N=1$ (this amounts to the 1 loop string beta function
calculation) the correction to the K\"ahler potential is given by
\begin{equation} 
\delta K^{1L}={1\over \epsilon } \ln (\mbox{\rm det}g),
\quad\mbox{with}\quad
g_{I\bar{J}}=\partial_I\partial_{\bar{J}}K(\Phi,\bar{\Phi}),
\end{equation}
where ($\epsilon=d-2$) we are using dimensional regularization. 
If we write this
in powers of the background field $\Phi$ we arrive at the expansion
shown diagrammaticaly in figure 3. 
\begin{figure}[h]
\hskip 120pt\includegraphics{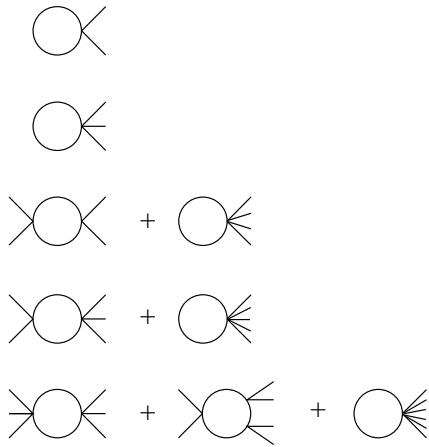}
\label{Fig3}\caption{1 loop expansion}
\end{figure}
Each vertex corresponds to a
term in the expansion of the K\"ahler potential. The external lines
correspond to the number of background fields. The first line thus
corresponds to the Ricci tensor 
$R_{I\bar{J}}=\delta^{K\bar{K}}K_{IK\bar{J}\bar{K}}$ 
evaluated at the special point about which
we have chosen the normal coordinates.\footnote{The series starts at
order $\Phi^2$ since this is the first relevant contribution inside
the full superspace integral.} The second line corresponds to a
correction of order $\Phi^3$ etc. Saying that the metric is Ricci flat at
the point $\Phi=0$ amounts to having the first term equal to
zero. Saying that it is Ricci flat everywhere implies that every line,
(the coefficient for each power of $\Phi$) is zero. 

Now we consider general $N$. The 1 loop contribution now reads 
\begin{equation} 
\delta K^{1L}={1\over \epsilon } \sum_{ij}\ln (\mbox{\rm det}(g_{ij})),
\quad\mbox{with}\quad
(g_{I\bar{J}})_{ij}={\partial^2\over\partial\Phi^I_{ij}
\partial\Phi^{\bar{J}}_{ji}}K(\Phi,\bar{\Phi}),
\end{equation}
where the determinant is taken over the indices $I$ and $\bar{J}$. As
discussed in section(4.2) this contribution has to be set to
zero. Equivalently this leads to  the condition that for each $i$, $j$
\begin{equation}
\mbox{\rm det}(g_{ij})=1.
\label{det1}
\end{equation}
This is precisely one of the mass conditions deduced in \cite{DOK}

This condition first becomes nontrivial for the third line of
figure(3) which represents the sum of a term coming from all
possible connected contractions of the 
two fourth order terms in the K\"ahler
potential and the contraction of single sixth order term with itself.

There were two mass conditions found in \cite{DOK}. Imposing them both
constrained the fourth order term to be the totally symmetrized
product
\begin{equation}
K^{(4)}=K_{I K \bar J \bar L} \bigl[
{2\over 3}\,\Phi^I \Phi^K \bar\Phi^{\bar J} \bar \Phi ^{\bar L}\,+\,
{1\over 3}\,\Phi^I \bar\Phi^{\bar J} \Phi^K \bar \Phi ^{\bar L} \bigr].
\label{symK4}
\end{equation}

For the sixth order term it was found in \cite{DOK} that the symmetrized
trace was no longer sufficient to satisfy the mass conditions, and an
explicit sixth order term which did satisfy them was
constructed. This term was somewhat complicated. Since in fact there
exists a very simple solution to the condition (\ref{det1}), we think
it worth presenting below.

The contraction of two symmetrized fourth order terms of the form 
(\ref{symK4}) gives rise to a contribution proportional to
\begin{eqnarray}
\delta K_{K_4K_4}
\,=\,
&\delta^{M\bar{M}}\delta^{P\bar{P}}
\bigl(K_{IM\bar{J}\bar{P}}K_{KP\bar{l}\bar{M}}
+{1\over 4}K_{IK\bar{M}\bar{P}}K_{MP\bar{J}\bar{L}}\bigr)
{16\over 9}{\cal P}(I,K,\bar{J},\bar{L})\cr
&\hskip 100pt+\delta^{M\bar{M}}\delta^{P\bar{P}}
K_{IM\bar{J}\bar{P}}K_{KP\bar{L}\bar{M}}
{8\over 3}{\cal Q}(I,K,\bar{J},\bar{L}),
\label{K4K4}
\end{eqnarray}
where ${\cal P}$ and ${\cal Q}$ are polynomials in the diagonal chiral
fields. They are given by
\begin{eqnarray}
{\cal P}(I,K,\bar{J},\bar{L})
=&\hskip -60pt(2N+7)\sum_iI_iK_i\bar{J}_i\bar{L}_i
\hskip 140pt\cr
&\hskip -30pt +\sum_{i\neq j}\bigl[
2I_iK_i\bar{J}_j\bar{L}_j+
{1\over 2}I_iK_j\bar{J}_i\bar{L}_j+{1\over 2}I_iK_j\bar{J}_j\bar{L}_i)\cr
&\hskip 50pt
+(I_jK_i\bar{J}_i\bar{L}_i+I_iK_j\bar{J}_i\bar{L}_i
+I_iK_i\bar{J}_j\bar{L}_i+I_iK_i\bar{J}_i\bar{L}_j)\bigr],
\label{Pdef}
\end{eqnarray}
\begin{eqnarray}
\hskip -170pt{\cal Q}(I,K,\bar{J},\bar{L})=\sum_{i\neq j}\bigl[
I_iK_i\bar{J}_j\bar{L}_j-I_iK_j\bar{J}_i\bar{L}_j\bigr]
\label{Qdef}
\end{eqnarray}
where for compactness we denote the chiral fields solely by their complex
indices, i.e. $I_i=\Phi^I_i$, $\bar{J}_i=\Phi^{\bar{J}}_i$ etc. We have
split the result up into two polynomials for reasons that will become
clear shortly. The two possible contractions give rise to polynomials
that are almost exactly equivalent. The first contraction in
(\ref{K4K4}) differs from the second in that the coefficients of two
of the terms in the second line of ${\cal P}$ are interchanged. The
addition of the polynomial ${\cal Q}$ serves to interchange these two
coefficients. 

The sixth order term in the K\"ahler potential has the general form
\begin{equation}
K_{IKM\bar{J}\bar{L}\bar{N}}\bigl(
\alpha\,IKM\bar{J}\bar{L}\bar{N}
+\beta_1\,IK\bar{J}M\bar{L}\bar{N}
+\beta_2\,IK\bar{J}\bar{L}M\bar{N}
+\gamma\,I\bar{J}K\bar{L}M\bar{N}\bigr),
\label{genK6}
\end{equation}
where for now we are neglecting possible commutator terms.
The coefficient $K_{IKM\bar{J}\bar{L}\bar{N}}$ is symmetric under
arbitrary interchange of holomorphic indices and under interchange of
antiholomorphic indices. Again we denote, for compactness,
$\Phi^I_i=I_i$ etc. The totally symmetrized product (within a trace) 
corresponds to  $\alpha=\beta_1=\beta_2=3/10$ and $\gamma=1/10$.

Ignoring the possible commutator terms we sum over all possible 
contractions of a holomorphic and
anti-holomorphic index in equation(\ref{genK6}) and look for
coefficients $\alpha$, $\beta_1$, $\beta_2$ and $\gamma$ for which
the resulting polynomial resembles as much as possible
equation(\ref{K4K4}).  We find that, for 
$\alpha=1$ and all other constants zero, 
\begin{equation}
\delta K_{K_6}
\,=\,
\delta^{M\bar{M}}
K_{IKM\bar{J}\bar{L}\bar{N}}\,{\cal P}(I,K,\bar{J},\bar{L}),
\label{delK6}
\end{equation}
where the polynomial ${\cal P}$ is defined in (\ref{Pdef}). To
complete the sixth order term we now add a
term which upon contraction will cancel with ${\cal Q}$ polynomial
contribution in equation(\ref{K4K4}). The term required is 
\begin{equation}
-{4\over 3}\delta^{P\bar{P}}K_{IM\bar{J}\bar{P}}K_{KP\bar{L}\bar{N}}
\bigl(MIK\bar{N}\bar{J}\bar{L}-MI\bar{J}\bar{N}K\bar{L}
+\bar{N}IKM\bar{J}\bar{L}-\bar{N}I\bar{J}MK\bar{L}\bigr).
\end{equation}
It is relatively easy to see that contracting in all possible ways a
holomorphic with an anti-holomorphic index does indeed give the term
proportional to ${\cal Q}$. Firstly contracting $K$ with $\bar{J}$ is
zero by construction and furthermore contracting $K$ with $\bar{L}$ or
$\bar{N}$ automatically gives zero since the result is proportional to
the Ricci tensor. One thus only needs to consider the contractions between
$IM$ and $\bar{L}\bar{N}$, which by symmetry reduces to contractions
between $M$ and $\bar{N}$.

The total sixth order term is thus
\begin{eqnarray}
K_6=&\hskip -150pt
K_{IKM\bar{J}\bar{L}\bar{N}}\,IKM\bar{J}\bar{L}\bar{N}\hskip 60pt\cr
&\hskip -50pt
-{4\over 3}\delta^{P\bar{P}}K_{IM\bar{J}\bar{P}}K_{KP\bar{L}\bar{N}}
\bigl(MIK\bar{N}\bar{J}\bar{L}-MI\bar{J}\bar{N}K\bar{L}\cr
&\hskip 170pt+\bar{N}IKM\bar{J}\bar{L}-\bar{N}I\bar{J}MK\bar{L}\bigr).
\label{K6t}
\end{eqnarray}
The only other possible terms that could be added to this are terms
which disappear under the contractions being considered above,
i.e. terms which are zero when there are less than three non diagonal
matrices. Such terms can only be constructed from the product of three
commutators. Presumably the difference
between the result (\ref{K6t}) and the complicated form presented in
\cite{DOK} amounts to the addition of such terms.

\section{ D0 Branes in a Curved Background}

The analysis for string matrix theory in a curved background can be
applied to the quantum mechanics of D0 branes. This is obtained by
further reducing the $N=4$ 4d SYM theory to 0+1 dimensions.  This
describes the evolution of D0 branes in a six dimension K\"ahlerian
background.  Using the background field method one can study the
effective action obtained from integrating out the off-diagonal
fields and the fluctuations of the diagonal part. The analysis is similar
to the calculations presented in the previous section. The
only difference being the dimension of the loop integrals; one
integrates over a single momentum variable.  All the loop integrals
are therefore UV finite.

Integrating out the background fields thus leads to non-local, (in
spacetime)  finite terms coupling two or more diagonal elements/D0
branes. As discussed in the previous section, by a judicious choice of
matrix ordering, it might be possible to ensure that all such terms
are zero for Ricci flat metrics at the one, two and three loop
level. At four loops however Ricci flatness is insufficient to cancel
the non-local terms and the manifold has to be further restricted to
have vanishing six dimensional Euler class. This is the case for a
hyper-K\"ahler surface.

\vskip 1 cm
\section {Conclusions}
We have shown that the 1 loop calculation for the effective action for
matrix string theory in a curved space has divergences corresponding
to non-local terms connecting together two or more diagonal
elements. These terms arise from simple matrix generalizations of the
string theory beta function calculation. They correspond to
powers of traces and, since the original
action is postulated to contain a single trace,
cannot be renormalized into a redefinition of the K\"ahler
potential. The condition that these terms vanish is identical to one
of the two mass conditions imposed on the K\"ahler potential in the
analysis of Douglas et. al. \cite{DOK}. At lowest nontrivial order
it is possible to find particular matrix orderings and commutator
terms that satisfy the condition. 

However the fact that the four loop term cannot be renormalized into
the K\"ahler potential means that these models have a limited
range of applicability, only being consistent for Ricci flat manifolds
with vanishing six dimensional Euler density. 

The analysis of this article did not depend on the size of the
matrices and it is hard to see any hidden subtleties in the taking of
the large $N$ limit that might change the analysis for infinite $N$.
\footnote{Some subtleties in the large $N$ limit have been pointed 
out in \cite{wyn2} but they are infra-red effects not ultra-violet.}

It is also not at all obvious how to modify the gauged matrix 
sigma models to have a more general applicability. The addition by
hand of powers of traces to cancel the divergences would be ad hoc and
it is not clear how the inclusion of higher derivative terms could
improve the problem. It seems likely that there is something more
fundamental missing from the description. Certainly one is all too
aware of the lack of a basic principle to guide us and the lack of a 
solid set
of fundamental building blocks from which to construct
actions. Perhaps this is another sign \cite{mos,do} that matrix variables
are insufficient to describe curved space, even for infinite $N$. 

\section {Appendix}
\vskip 1 cm

The superfield reduction of section 4 was based on the
background superfield method first devised in \cite{GSR} to discuss
properties of $N=4$ SYM theories. For the matrix string theory on a
curved background this method is crucial to integrate out the
background fields and obtain the effective action of the background
diagonal fields.  The idea is a simple superfield generalization of
the background field method used to quantize gauge theories. We refer 
the reader to
\cite{GSR}\cite{onethou}\cite{west}
for fuller details.

Recall that the gauge transformation of $N=1$ vector superfields are given by
\begin{equation}
e^V\to e^{\Lambda}e^Ve^{-\Lambda}
\end{equation}
where $\Lambda$ is a chiral superfield satisfying 
$\bar D_{\alpha}\Lambda=0$ i.e. the chiral superfields
are annihilated by the fermionic covariant derivatives.
Let us now assume that the vector superfield is the sum of 
a background configuration and a fluctuating part.
The most natural splitting would be to define the vector field 
as the sum of these two contributions. However the quantization 
of the theory with
a background is easier if one splits 
\begin {equation}
e^V=e^{V_B\over 2}e^ve^{-{V_B\over 2}}
\end{equation}
where $v$ is the fluctuating part and $V_B$ the background equation. 
To first order in the Campbell-Hausdorff expansion this leads to
\begin{equation}
V\sim V_B +v
\end{equation} 
as expected.
The chiral part of the Lagrangian is constructed using the 
background covariant derivatives
\begin{equation}
{\cal D}_{\alpha}= e^{-{V_B\over 2}}D_{\alpha}e^{V_B\over 2}
\label{calD}
\end {equation}
and its  conjugate.  
Indeed one defines background covariantly chiral superfield as the 
solutions of
\begin{equation}
\bar{\cal D}_{\alpha} \Phi=0
\end{equation}
This constraint is solved defining 
\begin{equation}
\Phi= e^{V_B\over 2}\tilde \Phi e^{-{V_B\over 2}}
\end{equation}
and $\tilde \Phi$ is a chiral superfield.

These ingredients can be used to write the Lagrangian in a way 
that naturally separates the background fields from the fluctuating parts.
The chiral superfield containing the field strengths of the gauge fields is
\begin{equation}
e^{V_B\over 2}W_{\alpha}e^{-{V_B\over 2}}
={i\over 2}[{\cal D}^{\beta}, 
\{e^{-v}{\cal D}_{\alpha}e^v,{\cal D}_{\beta}\}]
\end{equation}
The chiral part of the Lagrangian can be written in 
terms of the background chiral fields and the fluctuating 
part of the gauge fields.
For instance the canonical term becomes
\begin{equation}
\hbox{tr} (e^V\tilde \Phi e^{-V}\tilde\Phi^{\dagger})
=\hbox {tr} (e^v\Phi e^{-v} \bar \Phi ^{\dagger})
\end{equation}
Similarly the whole K\"ahler potential is a function of 
$v$, $\Phi$ and $\bar\Phi$ only. 
Likewise the superpotential becomes
a function $W(\Phi)$ of $\Phi$ only.
Finally the background chiral fields can be split into a  
background part  and a fluctuating part
\begin{equation}
\Phi=\Phi_B+\phi
\end{equation}
where both fields are background chiral.

The advantage of redefining the Lagrangian in such a way is 
that the effective action obtained  after integrating
over the fluctuations around the background field  is gauge invariant.  
The effective action is obtained after fixing the gauge. The gauge fixing term
is chosen to be
\begin{equation}
{\cal D}^2 V=\bar f, \ \overline{\cal D}^2 V=f
\end{equation}
where $f$ is a background chiral field.
This leads to the introduction of three ghosts, the two Fadeev-Popov ghosts 
and the Kallosh-Nielsen ghost. Using all these ingredients one can 
apply the usual procedure to calculate super-Feynman diagrams. 

The fact that the fields are covariantly chiral leads to more complicated 
propagators than is the case for ordinary chiral fields but for the divergent 
diagrams we consider in this article the distinction is unimportant, 
allowing us to use standard superfield Feynman rules.

\setlength{\baselineskip}{0.6666666667\baselineskip}

\end{document}